\documentclass[12pt]{article}
\usepackage{amsmath,amsfonts,amssymb}
\usepackage{citesort}
\usepackage{epsfig}

\def\bit{\begin{itemize}}
\def\eit{\end{itemize}}
\def\ben{\begin{enumerate}}
\def\een{\end{enumerate}}
\def\bed{\begin{description}}
\def\eed{\end{description}}
\def\susy{{\sc susy}}

\def\k{\kappa}
\def\l{\lambda}

\def\third{\frac{1}{3}\,}

\def\R{ {\rm R \kern -.31cm I \kern .15cm}}
\def\C{ {\rm C \kern -.15cm \vrule width.5pt \kern .12cm}}
\def\Z{ {\rm Z \kern -.27cm \angle \kern .02cm}}
\def\N{ {\rm N \kern -.26cm \vrule width.4pt \kern .10cm}}
\def\1{{\rm 1\mskip-4.5mu l} }
\def\lsim{\raise0.3ex\hbox{$<$\kern-0.75em\raise-1.1ex\hbox{$\sim$}}}
\def\gsim{\raise0.3ex\hbox{$>$\kern-0.75em\raise-1.1ex\hbox{$\sim$}}}
\def\noi{\noindent}

\def\beq{\begin{equation}}   
\def\eeq{\end{equation}}
\def\bea{\begin{eqnarray}}  
\def\eea{\end{eqnarray}}
\newcommand{\ba}{\begin{array}}
\newcommand{\ea}{\end{array}}
\def\nn{\nonumber}
\def\noi{\noindent}
\def\beeq{\begin{eqnarray}} \def\eeeq{\end{eqnarray}}
\newcommand\mysection{\setcounter{equation}{0}\section}
\renewcommand{\theequation}{\thesection.\arabic{equation}}
\newcounter{hran} \renewcommand{\thehran}{\thesection.\arabic{hran}}

\def\bmini{\setcounter{hran}{\value{equation}}
   \refstepcounter{hran}\setcounter{equation}{0}
   \renewcommand{\theequation}{\thehran\alph{equation}}\begin{eqnarray}}

\def\bminiG#1{\setcounter{hran}{\value{equation}}
\refstepcounter{hran}\setcounter{equation}{-1}
\renewcommand{\theequation}{\thehran\alph{equation}}
\refstepcounter{equation}\label{#1}\begin{eqnarray}}

\def\emini{\end{eqnarray}\relax\setcounter{equation}{\value{hran}}\renewcommand{\theequation}{\thesection.\arabic{equation}}}

\begin{document}
\centerline{\Large\bf NMHDECAY 2.0: An updated program}
\par \vskip 3 truemm
\centerline{\Large\bf for sparticle masses, Higgs masses,}
\par \vskip 3 truemm
\centerline{\Large\bf  couplings and decay widths in the NMSSM}
\vskip 1 truecm

\begin{center}
{\bf Ulrich Ellwanger}\footnote{E-mail :
ulrich.ellwanger@th.u-psud.fr},
{\bf Cyril Hugonie}\footnote{E-mail :
cyril.hugonie@th.u-psud.fr}
\par \vskip 5 truemm

Laboratoire de Physique Th\'eorique\footnote{Unit\'e
Mixte de Recherche - CNRS - UMR 8627}\par
  Universit\'e de Paris XI, B\^atiment
210, F-91405 Orsay Cedex, France\par \vskip 5 truemm

\end{center}
\vskip 1 truecm

\begin{abstract} 
We describe the the improved properties of the NMHDECAY program, that
is designed to compute Higgs and sparticle masses and Higgs decay
widths in the NMSSM. In the version 2.0, Higgs decays into squarks and
sleptons are included, accompagnied by a calculation of the squark,
gluino and slepton spectrum and tests against constraints from LEP and
the Tevatron. Further radiative corrections are included in the Higgs
mass calculation. A link to MicrOMEGAs allows to compute the dark
matter relic density, and a rough (lowest order) calculation of
BR$(b \to s \gamma)$ is performed. Finally, version 2.1 allows to
integrate the RGEs for the soft terms up to the GUT scale.
\end{abstract}

\vskip 1 cm

PAC numbers: 12.60.Jv, 14.80.Cp, 14.80.Ly

\vfill
\noi LPT Orsay 05-52\par

\noi July 2005 \par

\newpage \pagestyle{plain} \baselineskip 20pt

\mysection{Introduction}

\hspace*{\parindent}
The Next to Minimal Supersymmetric Standard Model
(NMSSM)~\cite{allg,bast,radcor1,radcor2,higsec1,higsec2,higsec3,yeg,higrad2}
provides a very elegant solution to the $\mu$ problem of the MSSM  via
the introduction of a singlet superfield $\widehat{S}$. For the
simplest possible scale invariant form of the superpotential, the
scalar component of $\widehat{S}$ acquires naturally a vacuum
expectation value of  the order of the \susy\ breaking scale, giving
rise to a value of $\mu$  of order the electroweak scale. Hence the
NMSSM is the simplest supersymmetric extension of the standard model in
which the fundamental Lagrangian contains just \susy\ breaking terms
but no other parameters of the order of the electroweak scale.\par

In addition, the NMSSM renders the ``little fine tuning problem'' of
the MSSM, originating from the non-observation of a neutral CP-even
Higgs boson at LEP II, less severe~\cite{bast}.\par

As in the MSSM, the phenomenology of the NMSSM depends on a certain
number of parameters (mostly soft \susy\ breaking parameters) that
cannot be predicted from an underlying theory at present. It is then
useful to have computer codes that compute physically relevant
quantities as Higgs and sparticle masses, couplings, decay widths etc.
as functions of the initial parameters in the Lagrangian. Such codes
allow to investigate which regions in parameter space are in conflict
with present constraints on physics beyond the standard model and, most
importantly, which regions in parameter space can be tested in future
experiments and/or astrophysical measurements.

In the MSSM, corresponding computer codes are HDECAY~\cite{hdecay},
FeynHiggs~\cite{feynhiggs}, Isajet~\cite{isajet}, 
SoftSusy~\cite{softsusy}, MicrOMEGAs~\cite{MicrOMEGAs}, 
Suspect~\cite{suspect}, Spheno~\cite{spheno}, 
CPSUPERH~\cite{cpsuperh}, SDECAY~\cite{sdecay} and
DARKSUSY~\cite{darksusy}.  
In the NMSSM, the only available code at present is
NMHDECAY~\cite{nmhdecay}. 

In the present paper we describe the improvements performed in the
version 2.0 of NMHDECAY. (In the meantime version 2.1 is available,
whose features are described in the file {\sf README} on the web page
\cite{nmhdecay}. In the cases where they differ from the ones of
version 2.0 they are described below.) First we recall the features of
the previous version of NMHDECAY, version 1.1. Starting from a set of
(low energy) parameters it performs the following tasks:

\bit

\item It computes the masses and couplings of all physical states
in the Higgs sector and in the chargino and neutralino sectors.

\item It computes the branching ratios into two particle final states
(quarks and leptons, all possible combinations of gauge and
Higgs bosons, char\-ginos, neutralinos, but not decays into squarks and
sleptons) of all 6 Higgs particles of the NMSSM. Three body decays via
$WW^*$ and $ZZ^*$ are computed as in HDECAY \cite{hdecay}, but no four
body decays are taken into account.

\item It checks whether the Higgs masses and couplings violate any
bounds from negative Higgs searches at LEP, including many quite
unconventional channels that are relevant for the NMSSM Higgs sector.
It also checks the bound on the invisible $Z$ width (possibly violated
for light neutralinos). In addition, NMHDECAY 1.1 checks the LEP bounds
on the lightest chargino and on neutralino pair production.

\item It checks whether the running Yukawa couplings $\lambda$, $\kappa$,
$h_t$ or $h_b$ encounter a Landau singularity below the GUT scale.

\item Finally, NMHDECAY 1.1 checks whether the physical minimum (with all
vevs non-zero) of the scalar potential is deeper than the local
unphysical minima with vanishing $\left< H_u \right>$, $\left< H_d
\right>$ or $\left< S \right>$.

\eit

The improvements in the versions 2.0+ are as follows:

\ben

\item Further radiative corrections are added in the Higgs sector, in
order to improve the precision of the Higgs masses calculations. 
In addition, all squark and slepton masses (and mixing angles for
the third generation) are computed. 

\item Branching ratios of all Higgs states into squarks and sleptons
are computed, and squark and slepton loops are included in the Higgs
decays to two gluons and two photons.

\item Experimental constraints from LEP and Tevatron on squark, gluino
and slepton masses are checked.

\item The dark matter relic density can be computed, via a call of a
NMSSM version of MicrOMEGAs (that is provided on the same web site).

\item The branching ratio $BR(b \to s \gamma)$ is computed to lowest
order.

\item In the version 2.1, the RGEs for the soft /susy/ breaking terms
can be integrated up to $M_{GUT}$.

\item A Makefile for optimal compilation is provided.

\een

The conventions concerning the Lagrangian of the model are the
same as in version~1.1: The superpotential $W$ is given by
\beq\label{1.1e}
W = h_t \widehat{Q}\cdot \widehat{H}_u \widehat{T}_R^c - h_b
\widehat{Q} \cdot  \widehat{H}_d  \widehat{B}_R^c - h_{\tau}
\widehat{L} \cdot \widehat{H}_d \widehat{L}_R^c +\l \widehat{S}
\widehat{H}_u \cdot \widehat{H}_d + \frac{1}{3} \k \widehat{S}^3 \ .
\eeq 
\noi (Hereafter, hatted capital letters denote superfields,
and unhatted capital letters the corresponding (complex) scalar
components.) The $SU(2)$ doublets are
\beq\label{1.2e}
\widehat{Q} = \left(\ba{c} \widehat{T}_L \\ \widehat{B}_L
\ea\right) , \
\widehat{L} = \left(\ba{c} \widehat{\nu}_{\tau L} \\ \widehat{\tau}_L
\ea\right) , \
\widehat{H}_u = \left(\ba{c} \widehat{H}_u^+ \\ \widehat{H}_u^0
\ea\right) , \
\widehat{H}_d = \left(\ba{c} \widehat{H}_d^0 \\ \widehat{H}_d^-
\ea\right) .
\eeq

\noi Products of two $SU(2)$ doublets are defined as, e.g.,
\beq\label{1.3e}
\widehat{H}_u \cdot \widehat{H}_d = \widehat{H}_u^+ \widehat{H}_d^- 
- \widehat{H}_u^0 \widehat{H}_d^0\ .
\eeq

\noi For the soft \susy\ breaking terms we take
\bea\label{1.4e}
-{\cal L}_\mathrm{soft} &=
m_\mathrm{H_u}^2 | H_u |^2 + m_\mathrm{H_d}^2 | H_d |^2 + 
m_\mathrm{S}^2 | S |^2 +m_Q^2|Q^2| + m_T^2|T_R^2| \nn \\
&+m_B^2|B_R^2| +m_L^2|L^2| +m_\mathrm{\tau}^2|L_R^2|
\nn \\
&+ (h_t A_t\ Q \cdot H_u T_R^c - h_b A_b\ Q \cdot H_d B_R^c - h_{\tau}
A_{\tau}\ L \cdot H_d L_R^c\nn \\
& +\l A_\l\ H_u \cdot H_d S + \third \k A_\kappa\ S^3 + \mathrm{h.c.}
)\,. \eea
The resulting mass matrices and couplings can be found in the appendix
of ref.~\cite{nmhdecay}. The conventions are also listed
within the FORTRAN code as comments
at the beginning of each corresponding subroutine.

The input parameters relevant for the Higgs sector of the NMSSM (at
tree level) are
\beq \label{1.5e}
\lambda,\  \kappa,\ A_{\lambda},\  A_{\kappa},\ \tan\!\beta\ 
=\ \left< H_u \right>/ \left< H_d \right>,\
\mu_\mathrm{eff}= \lambda \left< S \right>\ .
\eeq

As in the case of version 1.1, it is possible to a) use input and
output formats according to the \susy\ Les Houches Accord (SLHA)
conventions~\cite{slha} (with, however, a modified switch (65 instead
of 23) for $\mu_\mathrm{eff}$), b) use a privately defined
input and output format (scan), that allows to scan over a user defined
range of the input parameters (\ref{1.5e}).

In the next section, we describe the improvements of versions 2.0 and
2.1. Apart from the additional radiative corrections, we discuss the
precise meaning (renormalization scale) of the input parameters. In
section 3, we describe in detail how the different versions NMHDECAY
can be installed, compiled, and linked to MicrOMEGAs. We conclude with
a short outlook.

\mysection{Improvements in NMHDECAY 2.0 and 2.1}

\subsection{Radiative corrections to the Higgs masses}

Concerning radiative corrections induced by (s)top and (s)bottom loops,
we adopt the same philosophy as in version 1.1: First we compute the
running Yukawa couplings $h_t$, $h_b$ and $\lambda$ as well as the
Higgs vevs (and $\tan\!\beta$) at the scale of the stop and sbottom
masses, taking into account effects of order $h_{t/b}^2$ times
potentially large logarithms. (The input parameters $\lambda$ (and
hence $\mu_{\mathrm{eff}}$), $\kappa$, $A_\lambda$, $A_\kappa$ are
defined at the scale $M_Z^2$, which we can identify with $m_{top}^2$
within the present approximation.) Then we add the one loop radiative
corrections $\sim h_{t/b}^4$ to the effective potential, taking the
full dependence on stop/sbottom masses and mixing angles into account.
The advantage of performing this calculation at the scale of the
stop/sbottom masses is that the remaining dominant two loop corrections
$\sim h_{t/b}^6$, $\sim h_{t/b}^4\alpha_s$ are relatively simple. These
latter corrections include now also the dependence on $h_b$ (in
contrast to the version 1.1) in order to cover the large $\tan\!\beta$
regime. Also, as in version 1.1, due to this procedure several non
leading two loop effects (related to squark mass splittings and
mixings) are automatically included, as in ref.~\cite{car1} for the
MSSM.

New in the version 2.0 are the corrections of the order $\sim g^2
h_{t/b}^2$ to the CP even Higgs boson masses (where $g$ denotes the
electroweak gauge couplings), induced by the stop/sbottom D term
couplings, whose dependence on the stop/sbottom masses is
computed beyond the leading logarithmic approximation.

Concerning the logarithmic one loop corrections of the order $\sim g^4$
(to the mass of the lighter doublet like CP even state), we distinguish
now the masses of the different squarks/sleptons of the different
generations (assuming the first two generations to be degenerate). New
in the version 2.0 are the logarithmic one loop corrections to fourth
order in the NMSSM specific Yukawa couplings $\lambda$ and
$\kappa$, that have only recently been computed in ref.~\cite{yuk}.

Finally, in version 2.0 the corrections of the order $\sim g^2
h_{t/b}^2$ to the Higgs pole masses (from contributions $\sim
h^2_{t/b}$ to the Higgs self energies as in ref. \cite{pierce}) are
included.

The dominant sources of uncertainty on the mass of the lighter doublet
like neutral Higgs boson mass are thus non logarithmic contributions of
the orders $g^4$, $g^2\lambda^2$, $\lambda^4$, $g^2\kappa^2$,
$\kappa^4$ and $\lambda^2\kappa^2$, and two loop contributions beyond
the dominant double  logarithms $\sim h_{t/b}^6, h_{t/b}^4\alpha_s$.

\subsection{Higgs decays}

In addition to the Higgs decays considered in the version 1.1, the
version 2.0 of NMHDECAY includes now the possible two body decays of
all CP even, CP odd and charged Higgs bosons into all squarks and
sleptons. Also, squark and slepton loop contributions to the
radiatively induced decays of the CP even and CP odd Higgs bosons into
two photons and two gluons are included. 

The corresponding code is essentially as in HDECAY~\cite{hdecay}, but
with the more complicated Higgs self couplings and mixings in the
NMSSM. The leading logarithmic corrections from top and bottom quark
loops to the Higgs self couplings are included, and  the Higgs squark
couplings are scaled up to a scale $Q^2$ corresponding to the squark
masses which takes care of large logarithmic radiative corrections
$\sim h_{t/b}^2\ln(M_{Squark}^2/M_Z^2)$, neglecting terms $\sim
h_{t/b}^2\ln(M_{Higgs}^2/M_{Squark}^2)$.

\subsection{Sparticle masses}

The masses and mixing angles of the two charginos and five (in the
NMSSM) neutralinos were already calculated in the version 1.1 of
NMHDECAY in the subroutines CHARGINO and NEUTRALINO. In the
actual version 2.1, one loop radiative corrections to the neutralino
and chargino mass matrices are included as in section 4.2 in
\cite{pierce}.

In the subroutine MSFERM in the version 2.0+, also the slepton masses
are computed, and the squark mixing angles and pole masses are
calculated including the one loop $\alpha_s$ corrections\footnote{We
thank S. Kraml for contributions to the corresponding codes}.

The subroutine GLUINO in the versions 2.0+ includes a
computation of the gluino pole mass to the order $\alpha_s$.

In the actual version 2.1 we assume that the input values of the soft
\susy\ breaking terms are given at a \susy\ scale $Q^2 = (2
M_Q^2+M_U^2+M_D^2)/4$, where $M_Q$, $M_U$ and $M_D$ are the running
squark masses of the first two generations. If desired, this scale can
also be set by the user.

Thus, in the versions 2.0+ of NMHDECAY the complete sparticle
spectrum is computed. Already in the version 1.1, the masses of the two
charginos and the masses and couplings to the $Z$ boson of the five
neutralinos were compared to LEP constraints from direct searches and
constraints on the invisible $Z$ width. Now, in addition, NMHDECAY
tests the squark and gluino masses against constraints from the
Tevatron~\cite{sqgtev} and LEP~\cite{sqglep}. (The Tevatron constraints
are those used by the LEPSUSY Working group~\cite{sqglep}.) 
As usual, NMHDECAY issues a warning in case where any of the present
constraints is violated. NMHDECAY is thus quite unique in testing the
complete Higgs and sparticle spectrum against constraints from
accelerator experiments.

\subsection{Dark matter relic density}

The dark matter relic density in the NMSSM has recently been studied
in ref.~\cite{darkmatter1} (for previous investigations, see
refs.~\cite{darkmatter2}). For this purpose, NMHDECAY 1.1 was used
to compute the Higgs and sparticle spectrum, which was then fed into
a new version of MicrOMEGAs extended to the NMSSM. MicrOMEGAs
calculates all the relevant cross-sections for the lightest neutralino
annihilation and coannihilation. It then solves the density evolution
equation numerically, without using the freeze-out approximation, and
computes the relic density of the lightest neutralino.

In the version 2.0 of NMHDECAY, the dark matter relic density can be
computed for any choice of input parameters, by setting a simple flag
in the input file, through a link to this NMSSM version of MicrOMEGAs. 
The details on how this link has to be installed will be given in the
next section.

In case the corresponding flag is on, the computed amount of dark
matter is compared to constraints from WMAP~\cite{dmconstr} ($.0945\ 
<\ \Omega h^2\ <\ .1287$), and a warning is issued in case the
result is too large or too small. (Clearly, a warning corresponding to a
too small LSP relic density can be ignored, if one is ready to assume
additional contributions to the dark matter.)

\subsection{BR($b \to s \gamma$)}

In the version 2.0 of NMHDECAY the branching ratio BR($b \to s \gamma$)
is computed to lowest order in the subroutine BSG. Contributions from
charged Higgs and chargino/squark loops are included, which are the
same as in the MSSM. The theoretical error is about 6\% for $\tan\beta\
\lsim\ 3$, but increases rapidly with increasing $\tan\beta$. Hence the
result is only a rough estimate for $\tan\beta\ \gsim\ 5$.

\subsection{The RGEs}

A new feature of version 2.1 is the (possible) evolution of all NMSSM
parameters up to the GUT scale. Three different subroutines are
present:

\bit

\item the subroutine {\sf RGES} integrates the RGEs for the gauge and
Yukawa couplings up to the GUT scale (defined by the matching of the
$U(1)_Y$ and $SU(2)$ gauge couplings $g_1$ and $g_2$). It is always
called and serves to compute $M_{GUT}$ as well as the Yukawa couplings
at the GUT scale in order to check the absence of a Landau singularity
(by requiring any Yukawa coupling to be less than $4\pi$). Two loop
$\beta$ functions are used,
and the conversion from the ${\overline{MS}}$ to the ${\overline{DR}}$
scheme as well as all possible threshold effects between $M_Z$ and
the \susy\ scale are taken into account.

\item the subroutine {\sf RGESOFT} integrates the RGEs for the 
couplings and the soft terms up to the GUT scale (that is now known
from a previous call of {\sf RGES}). This subroutine is always called
in the version {\sf SLHA}, but in the version {\sf SCAN} it is called
only if the output format ``long'' is chosen. (It is not useful, but
time consuming, for large scans.) At present, terms $\sim \lambda^2$,
$\kappa^2$ in the two  loop coefficients of the $\beta$ functions for
the soft terms are missing (this will be improved in the near future). 

\item the subroutine {\sf RGESOFTINV} integrates the RGEs for the 
couplings and the soft terms from the GUT scale down to the \susy\
scale Q. It is called only if {\sf RGESOFT} is called. Unless modified,
it uses the outputs of the subroutines {\sf RGES} and {\sf RGESOFT} for
the values of all parameters at the GUT scale and leads thus to no new
results for the parameters at the \susy\ scale. Unless modified, it
produces no output and acts just as a dummy. However, the user can
easily modify the values of the (or some) parameters at the GUT scale
(by choosing {\sf INGUT=1}, after which the subroutine has to be
re-compiled), and generate an output for the resulting parameters at
the \susy\ scale. These can subsequently be used as input, and allow to
generate sets of parameters that have desirable properties (as partial
universality) at the GUT scale. Clearly, the Higgs masses squared at
the \susy\ scale cannot be put in directly, and can be modified only
indirectly by varying other parameters like $\tan\beta$,
$\mu_\mathrm{eff}$, $\lambda$, $\kappa$, $A_\lambda$ or $A_\kappa$. 

\eit

\mysection{How to install NMHDECAY}

Two versions of NMHDECAY are available:

\ben

\item nmhdecay\_slha uses an input file and produces output files that
are suitable generalizations of the SLHA
conventions~\cite{slha}. It is configured for studying the
properties of one user-defined point in parameter space.

\item nmhdecay\_scan employs privately defined input and  output files.
It allows to scan over parts of or all of the NMSSM parameters
$\lambda$, $\kappa$, $\mu_\mathrm{eff} = \lambda \left< S \right >$,
$\tan\!\beta$, $A_\lambda$ and $A_\kappa$.

\een

\noi Both programs are based on one single Fortran code ({\sf
nmhdecay\_slha.f} or {\sf nmhdecay\_scan.f}) contained in the
compressed directory {\sf NMHDECAY.tgz} that can be downloaded from the
web page {\sf{http://www.th.u-psud.fr/NMHDECAY/nmhdecay.html}}.

The directory {\sf NMHDECAY} contains also a {\sf Makefile} as well as
input files for both versions and test output files. Once the
compressed tar file {\sf NMHDECAY.tgz} is downloaded, one should type:

\begin{verbatim}
tar -zxvf NMHDECAY.tgz
cd NMHDECAY
./make
\end{verbatim}

\noi The command {\sf make} will create 2 executable files, {\sf
nmhdecay\_slha} and {\sf nmhdecay\_scan}. If one wishes to compile just
one of them, it suffices to type {\tt ./make nmhdecay\_slha} or {\tt
./make nmhdecay\_scan}. Both codes need data files in order to check
Higgs, sfermion and gluino mass bounds from LEP and Tevatron. These
data files are contained in the compressed tar file {\sf EXPCON.tgz}
that can be downloaded from the same web page. To uncompress it, simply type

\begin{verbatim}
tar -zxvf EXPCON.tgz
\end{verbatim}

\noi which generates a directory {\sf EXPCON}. By default
in NMHDECAY, the path to
the files containing the constraints is {\sf ../EXPCON} so that the
directory {\sf EXPCON} should be located at the same place as the
directory {\sf NMHDECAY}. However, if one wishes to place it somewhere
else, it is possible to define a corresponding environment variable
{\sf EXPCON\_PATH}:

\ben

\item[a)] From a C-Shell, type \\
{\tt setenv EXPCON\_PATH full\_path\_to\_EXPCON} \\
For a permanent setting add the same line
to the file {\sf .cshrc}.

\item[b)] From a Bash Shell, type\\
{\tt export EXPCON\_PATH=full\_path\_to\_EXPCON} \\
For a permanent setting add
the same line to the file {\sf .bashrc}.

\een

\noi Finally, for those wishing to compute the dark matter relic
density in the NMSSM, a version of MicrOMEGAs\_1.3 extended to the
NMSSM~\cite{darkmatter1} is contained in the compressed tar file {\sf
micromegas\_1.3\_nmssm.tgz}, also available on our web page.
In order to uncompress and to compile the
C code {\sf omg.c} and the MicrOMEGAs library, one should type:

\begin{verbatim}
tar -zxvf micromegas_1.3_nmssm.tgz
cd micromegas_1.3_nmssm
./micro_make omg.c
\end{verbatim}

\noi This will generate the executable file {\sf omg} that can be
called from NMHDECAY. It is possible to put the directory {\sf
micromegas\_1.3\_nmssm} wherever one wants, provided one defines the
environment variable {\sf MICROMG\_PATH} as above. Otherwise {\sf
micromegas\_1.3\_nmssm} has to be put in the same directory that
contains {\sf NMHDECAY} and {\sf EXPCON} ({\it ie} the default value
for {\sf MICROMG\_PATH} is {\sf ../micromegas\_1.3\_nmssm}). However,
if one wants to move the directory {\sf micromegas\_1.3\_nmssm} (and
consequently change the variable {\sf MICROMG\_PATH}) after the
MicrOMEGAs library has been compiled, one needs to recompile it. In
order to do this, one has first to remove the generated files using the
script {\sf clean} (included in the directory {\sf
micromegas\_1.3\_nmssm}). The sequence of commands is then:

\begin{verbatim}
mv  micromegas_1.3_nmssm new_path
setenv MICROMG_PATH new_path (C_Shell)
  or
export MICROMG_PATH=new_path (Bash-Shell)
cd new_path
./clean
./micro_make omg.c
\end{verbatim}

We now outline the particular features of the two versions of NMHDECAY.

\subsection{NMHDECAY\_SLHA}

The program {\sf nmhdecay\_slha} uses the input file {\sf slhainp.dat},
a version of which ({\sf slhainp.dat.test})
is contained in the directory {\sf NMHDECAY}. This
sample file appears in Table 1 below. Several comments on its contents
are in order: \\

\noi a) {\sf BLOCK MODSEL} contains the switch 3 (corresponding to the
choice of the model) with value 1, as attributed to the NMSSM in
ref.~\cite{slha}. Any other choice would cause the program to stop. We
have also added the switch 9 corresponding to a flag for the call to
MicrOMEGAs: if the flag is set to 0, the relic density is not
calculated (and MicrOMEGAs does not need to be compiled). 
If the flag
is set to 1, {\sf nmhdecay\_slha} calls the executable file {\sf omg}
which returns the relic density of the lightest neutralino if it is the
LSP. \\

\noi b) {\sf BLOCK SMINPUTS} contains important Standard Model
parameters.

\ben

\item The first is the inverse electromagnetic coupling constant
$\alpha_\mathrm{em}^{-1}$, now at the scale $Q=M_Z$
(in contrast to the version 1.1 where it was taken at the scale $Q=0$).

\item Second, since Higgs vevs and couplings are defined in terms of
$M_Z$, $M_W$ and $G_F$, an on shell scheme is used implicitly, and the
values of $M_Z$, $M_W$ and $G_F$ are required. $M_Z$, $G_F$ (and
$\alpha_s(M_Z)$) are defined in {\sf BLOCK SMINPUTS}, whereas the
numerical value of $M_W$ is defined in the subroutine {\sf INITIALIZE}.
This subroutine assigns default values to {\it all} parameters and
reads the files with the experimental constraints (contained in the
directory {\sf EXPCON}).

\item Third, as part of this block the running $b$ quark mass
$m_b(m_b)$, the top quark pole mass and $m_\tau$ are read in.

\item In addition, {\sf nmhdecay\_slha} needs the strange quark mass
$m_s(\overline{MS}$) and the charmed quark (pole) mass (taken as
$190$~MeV (at Q = 1 GeV) and $1.42$~GeV, respectively, as in
HDECAY~\cite{hdecay}), as well as the CKM matrix elements $V_{us}$,
$V_{cb}$ and $V_{ub}$. The numerical values of these five parameters
are defined in the subroutine {\sf INITIALIZE}. (For convenience, they
are printed out in the output file {\sf spectr.dat}, see below.)

\een

\noi c) {\sf BLOCK MINPAR} contains the optional switch 0 with the
input value for the renormalization scale Q at which the squark and
slepton masses, trilinear couplings and gaugino masses are defined. For
Q = 0 (or no switch 0) this renormalization scale is computed
internally from the average of the first generation squark masses. The
switch 3 corresponds to the input value for $\tan\!\beta$.\\

\noi d) {\sf BLOCK EXTPAR} contains the \susy\ and soft-\susy-breaking
parameters. An extention of the SLHA conventions is needed here. The
new entries are:

\begin{center}
\begin{tabular}{lll}
61 & \hspace*{.5in} & for $\lambda$ \\
62 & \hspace*{.5in} & for $\kappa$ \\
63 & \hspace*{.5in} & for $A_\lambda$ \\
64 & \hspace*{.5in} & for $A_\kappa$ \\
65 & \hspace*{.5in} & for $\mu_\mathrm{eff} = \lambda \left< S \right>$
\end{tabular}
\end{center}

\noi Note that in NMHDECAY 1.0, the switch 23 (corresponding to the
MSSM $\mu$ parameter) was used for $\mu_\mathrm{eff}$. Introducing a
new switch for this parameter allows, in principle, non zero values for
both $\mu$ and $\mu_\mathrm{eff}$, although such a scenario is not
implemented in NMHDECAY (the main motivation for the NMSSM is to get
rid of the MSSM $\mu$ parameter). A non zero value for $\mu$ in switch
23 is simply ignored. \\

The output files of {\sf nmhdecay\_slha} are {\sf spectr.dat}, {\sf
decay.dat} and {\sf omega.dat}. The directory {\sf NMHDECAY} contains
test versions of these files -- {\sf spectr.dat.test}
(see Table 2 below), {\sf decay.dat.test} and {\sf omega.dat.test} --
corresponding to the sample input file {\sf slhainp.dat}. The content
of {\sf spectr.dat} is as follows:\\

\noi a) {\sf BLOCK SPINFO} is followed by warnings (switch 3) if any
phenomenological constraint is violated. This segment of the output
also displays error messages (switch 4) if any of the Higgs, squark or
slepton states have a negative mass squared. No spectrum output is
produced in this latter case. \\

\noi b) {\sf BLOCK SMINPUTS} contains a printout of the Standard Model
input parameters. The numerical values for $M_W$, $m_s$, $m_c$,
$V_{us}$, $V_{cb}$ and $V_{ub}$, that have no SLHA numbers, also appear
in lines subsequent to {\sf \# SMINPUTS Beyond SLHA}. \\

\noi c) {\sf BLOCK MINPAR} is followed by a printout of the value of
$\tan\!\beta$. \\

\noi d) {\sf BLOCK EXTPAR} displays a printout of the \susy\ and
soft-\susy-breaking parameters. \\

\noi e) {\sf BLOCK MASS} contains the masses of all Higgs and sparticle
states. There, one finds several essential NMSSM generalizations of the
SLHA conventions. The new entries, with proposed PDG codes, are

\begin{center}
\begin{tabular}{lll}
45 & \hspace*{1in} & for the third CP-even Higgs boson, \\
46 & \hspace*{1in}  & for the second CP-odd Higgs boson, \\
1000045 &\hspace*{1in} & for the fifth neutralino. \\
\end{tabular}
\end{center}

\noi f) {\sf BLOCK LOWEN} contains observables relevant for precision
experiments at low energy, at present only the computed value for
BR$(b\ \to\ s \gamma)$ (switch 1). \\

\noi g) The Higgs mixings in the CP-even sector follow {\sf BLOCK
NMHMIX} and those in the CP-odd sector follow {\sf BLOCK NMAMIX}. Both
segments are required in order to parameterize the mixing in the
enlarged Higgs sector in the NMSSM. The meaning of the matrix elements
$S_{ij}$ (i, j = 1, 2, 3) and $P_{ij}$ (i, j = 1, 2, 3) is as follows.

\bit

\item According to the SLHA conventions the Higgs weak eigenstates
$H_u$, $H_d$ are denoted by $H_2$, $H_1$, respectively. (Inside the
Fortran code the Higgs states $H_u$, $H_d$ are denoted by $H_1$, $H_2$,
which is of no relevance for the SLHA output.)  Hence, for the purpose
of the SLHA output, the CP-even Higgs states are numbered by
$S^{weak}_i = (H_{dR}, H_{uR}, S_R)$ ($R$ refers to the real component
of the field). If  $h_i$ are the mass eigenstates (ordered in mass),
the convention is $h_i = S_{ij} S^{weak}_j$.

\item In the CP-odd sector the weak eigenstates are $H_{uI}, H_{dI},
S_I$ ($I$ for imaginary component). Again, for the purpose of the SLHA
output (NEW according to SLHA2 \cite{slha}!), the CP-odd Higgs states
are denoted by $H_{uI} = H_{2I}$, $H_{dI} = H_{1I}$ and $S_I =
H_{3I}$.  The mass eigenstates are $a_i$ where $a_1$ is the  Goldstone
mode $\tilde G$, and the two physical states $a_2$, $a_3$ are ordered
in mass. Then the elements of $P_{ij}$ are defined as $a_i = P_{ij}
H_{jI}$.


\eit

\noi h) {\sf BLOCK STOPMIX, SBOTMIX} and {\sf STAUMIX} contain the
mixing matrices of the stop squarks, sbottom squarks and stau sleptons
respectively, as defined in the MSSM. \\

\noi i) {\sf BLOCK NMNMIX} is followed by a printout of the obvious
generalization of the $4 \times 4$ MSSM neutralino mixing matrix to the
$5 \times 5$ NMSSM neutralino mixing matrix (with real entries); {\sf
BLOCK UMIX} and {\sf BLOCK VMIX} are followed by printouts of the $U$
and $V$ matrices as defined in the MSSM. \\

\noi h) As indicated, the subsequent {\sf BLOCKs GAUGE, YU, YD, YE,
L/K, AU, AD, AE, AL/AK} and {\sf MSOFT} contain the couplings and soft
\susy\ breaking terms at the \susy\ breaking scale, and the following
{\sf BLOCKs} with suffix {\sf GUT} the corresponding parameters at the
GUT scale.\\

The output file {\sf decay.dat} gives the decay widths of all Higgs
states into two particles, using the SLHA conventions and the above
generalizations of the PDG codes both for the decaying particle and the
final state particles. {\sf BLOCK DCINFO} gives informations about the
decay package (NMHDECAY version 2.1). \\

The output file {\sf omega.dat} contains the relic density of the
lightest neutralino (if {\sf OMGFLAG}=1). {\sf BLOCK RDINFO} gives
informations about the relic density package (MicrOMEGAs version 1.3).
{\sf BLOCK RELDEN} is then followed by switch 1 corresponding to the
relic density $\Omega h^2$ and switch 2 corresponding to a warning in
case the relic density of dark matter could not be computed or is
excluded by WMAP bounds.

\subsection{NMHDECAY\_SCAN}

The program {\sf nmhdecay\_scan} uses the input file {\sf scaninp.dat},
a version of which ({\sf scaninp.dat.test})
is downloaded automatically with the Fortran code
(see Table 3 below). In this input file, the following parameters must
be specified:

\bit

\item the total number of points to be scanned in parameter space;

\item the output format flag is 0 for ``short'', corresponding to
simple rows of numbers per allowed point in parameter space, and 1 for
``long'', as described below;

\item the flag {\sf OMGFLAG} (0 for no relic density computation, 1 for
relic density computation using MicrOMEGAs);

\item lower and upper limits for the NMSSM parameters $\lambda$,
$\kappa$, $\tan\!\beta$, $\mu_\mathrm{eff}$, $A_{\lambda}$ and
$A_{\kappa}$;

\item the soft squark and slepton masses, trilinear couplings and
the gaugino masses over all of which no scan is performed.

\eit

The scan in parameter space uses a random number generator, such that
all NMSSM parameters are randomly chosen point by point in the
parameter space within the specified limits. The standard model
parameters ($\alpha_s(M_Z)$, $G_F$, $\alpha_\mathrm{em}^{-1}$,
the lepton masses $m_\tau$ and $m_\mu$, the gauge boson masses $M_Z$
and $M_W$, the quark pole masses $m_s$, $m_c$, and $m_t$, the running
bottom quark mass $m_b(m_b)$ and the CKM matrix elements $V_{us}$,
$V_{cb}$ and $V_{ub}$) are specified in the subroutine {\sf
INITIALIZE}. \\

The output file containing the physical parameters is always called
{\sf scanout.dat}, regardless of  the output format chosen. The numbers
printed out for the output format 0 (recommended for scans over more
than ~10 points in parameter space) should be edited according to the
user's needs (see the section of the program in the subroutine
{\sf OUTPUT} following the comment line {\sf The following output can
be edited according to the user's needs}). The output format 1 is
easily readable and shows

\bit

\item the NMSSM parameters for each point as used as input, at the
\susy\ breaking scale, and at the GUT scale;

\item possible warnings in case any phenomenological constraint is
violated, or error messages ("fatal" errors) in case any of the Higgs
or sfermion states has a negative mass squared (in which case no
additional output is produced);

\item for each of the six Higgs states, their mass, their decomposition
in the basis of interaction eigenstates $(H_u,H_d,S)$, their reduced
couplings to gauge bosons (CV), up type quarks (CU), down type quarks
(CD), two gluons (CG) and two photons (CGA) (all relative to a standard
model Higgs boson with the same mass), their branching ratios (where
"Higgses" denote all possible two Higgs final states, and "sparticles"
all possible two particle neutralino/chargino/sfermion final states),
and their total width;

\item the neutralino, chargino and gluino masses (all masses in GeV),
as well as the neutralino composition in the basis  $\psi^0 = (-i\l_1 ,
-i\l_2, \psi_u^0, \psi_d^0,\psi_s)$;

\item the pole masses and cosine of the mixing angle for the stop and
sbottom states, as well as the pole masses of the squarks of the
first two generations;

\item the masses and cosine of the mixing angle for the stau states, as
well as the mass of the tau sneutrino and the masses of the first two
generation sleptons;

\item the dark matter relic density (if {\sf OMGFLAG=1});

\item the branching ratio BR$(b\ \to\ s\gamma)$.

\eit

The output file {\sf scanerr.dat} shows how many of the points in
parameter space have avoided fatal errors or violations of
phenomenological constraints, and the range in the NMSSM parameter
space over which points have passed all these tests. \\

Users who wish to call a subroutine as a function of the Higgs or
sparticle properties (masses, mixing angles and other quantities
computed during the course of the scan) should use the parameters
and common blocks found in the subroutine {\sf OUTPUT}.
The comments at the beginning of the main program should
allow easy identification of all the parameters, branching ratios,
mixing angles and so forth that would be of potential interest for
inputting into a user's subroutine.

\mysection{Summary and outlook}

The versions 2.0+ of NMHDECAY are both a calculator of the NMSSM Higgs
and sparticle spectrum, and a Higgs decay package. A
quite unique feature of NMHDECAY is the check of each point in
parameter space against limits on Higgs bosons and sparticles
from accelerator experiments (LEP and Tevatron), that would
have been impossible without the help of numerous collegues
participating in these experiments.

Apart from additional radiative corrections to the Higgs masses, the
new features of the version 2.0 of NMHDECAY are the inclusion of Higgs
decays into squarks and sleptons, the computation of the squark, gluino
and slepton masses  and mixings, the computation of the dark matter
relic density and the branching ratio BR$(b \to s \gamma)$.

Whereas the computation of the dark matter relic density is quite
reliable (as in the version 1.3 of MicrOMEGAs), the result for the
branching ratio BR$(b \to s \gamma)$ has still to be interpreted with
care, notably for larger ($\gsim\ 5$) values of $\tan\beta$.

It is clear that further improvements of NMHDECAY would be desirable:
more higher order corrections to both the Higgs masses and decay widths
would be welcome, with the aim to reach the present accuracy in the
MSSM. (Tests of NMHDECAY in the MSSM limit $\lambda, \kappa \to 0$ with
$\mu_{\mathrm{eff}}$ fixed indicate, however, that the deviation of the
mass of the lightest CP even Higgs boson w.r.t. corresponding MSSM
calculations, for the same CP odd Higgs pole mass and sparticle
spectra, is limited to about 3\% and mostly much smaller.)

Informations on low energy precision observables are included only in
the form of a rough calculation of BR$(b \to s \gamma)$, which should
certainly be improved. Also the anomalous magnetic moment of the muon as
well as $\Delta \rho$ should be computed.

We plan to treat these issues in the near future.

Finally, it would be useful to be able to choose (universal) 
soft \susy\ breaking terms at a {\sf GUT} scale as in a mSUGRA version
of the NMSSM. A corresponding code NMSPEC is in preparation.

\section*{Acknowledgments}

We thank P. Skands for comments on new SLHA and PDG particle codes, S.
Kraml for helpful contributions to the squark, slepton and gluino
sector, G. Belanger and A. Pukhov for help with the link to MicrOMEGAs, and M.
Schumacher, S. Hesselbach and W. Porod for comments on the previous version
of NMHDECAY.

\begin{table}[p]
{\footnotesize
\begin{verbatim}  
#  INPUT FILE FOR NMHDECAY VERSION 2.1
#  BASED ON SUSY LES HOUCHES ACCORD II
BLOCK MODSEL
    3     1         # NMSSM PARTICLE CONTENT
    9     0         # FLAG FOR CALL OF MICROMEGAS (0=NO, 1=YES)
BLOCK SMINPUTS
     1     1.27920000E+02   # ALPHA_EM^-1(MZ)
     2     1.16639000E-05   # GF
     3     1.17200000E-01   # ALPHA_S(MZ)
     4     9.11870000E+01   # MZ
     5     4.21400000E+00   # MB(MB)
     6     1.75000000E+02   # MTOP (POLE MASS)
     7     1.77710000E+00   # MTAU
BLOCK MINPAR
     0     4.68008177E+02   # REN. SCALE
     3     3.00000000E+00   # TANBETA
BLOCK EXTPAR
     1     8.16527293E+01   # M1
     2     1.53892994E+02   # M2
     3     4.72879088E+02   # M3
    11    -3.96388267E+02   # ATOP
    12    -7.44522774E+02   # ABOT
    13    -3.22488355E+02   # ATAU
    65     3.53337956E+02   # MU AT MZ
    31     2.40063601E+02   # M_eL
    32     2.40063601E+02   # M_muL
    33     2.39934370E+02   # M_tauL
    34     2.13138181E+02   # M_eR
    35     2.13138181E+02   # M_muR
    36     2.12838756E+02   # M_tauR
    41     4.74810720E+02   # M_q1L
    42     4.74810720E+02   # M_q2L
    43     4.20057207E+02   # M_q3L
    44     4.61771920E+02   # M_uR
    45     4.61771920E+02   # M_cR
    46     3.36293030E+02   # M_tR
    47     4.60437692E+02   # M_dR
    48     4.60437692E+02   # M_sR
    49     4.60051305E+02   # M_bR
    61     2.00000000E-01   # LAMBDA AT MZ
    62     1.47434810E-01   # KAPPA AT MZ
    63    -7.50350192E+01   # A_LAMBDA AT MZ
    64    -2.18338774E+00   # A_KAPPA AT MZ
\end{verbatim}}
\caption{Sample {\sf slhainp.dat} file.
\label{slhainp}}
\end{table}

\begin{table}
{\footnotesize
\baselineskip 10pt
\begin{verbatim}
# NMHDECAY OUTPUT IN SLHA FORMAT
# Info about spectrum calculator
BLOCK SPINFO        # Program information
     1   NMHDECAY   # spectrum calculator
     2   2.1        # version number
# Input parameters
BLOCK MODSEL
    3     1         # NMSSM PARTICLE CONTENT
BLOCK SMINPUTS
     1     1.27920000E+02   # ALPHA_EM^-1(MZ)
     2     1.16639000E-05   # GF
     3     1.17200000E-01   # ALPHA_S(MZ)
     4     9.11870000E+01   # MZ
     5     4.21400000E+00   # MB(MB)
     6     1.75000000E+02   # MTOP (POLE MASS)
     7     1.77710000E+00   # MTAU
# SMINPUTS Beyond SLHA:
# MW:     0.80420000E+02
# MS:     0.19000000E+00
# MC:     0.14000000E+01
# VUS:     0.22000000E+00
# VCB:     0.40000000E-01
# VUB:     0.40000000E-02
BLOCK MINPAR
     3     3.00000000E+00   # TANBETA
BLOCK EXTPAR
     1     8.16527293E+01   # M1
     2     1.53892994E+02   # M2
     3     4.72879088E+02   # M3
    11    -3.96388267E+02   # ATOP
    12    -7.44522774E+02   # ABOTTOM
    13    -3.22488355E+02   # ATAU
    65     3.53337956E+02   # MU
    31     2.40063601E+02   # LEFT SELECTRON
    32     2.40063601E+02   # LEFT SMUON
    33     2.39934370E+02   # LEFT STAU
    34     2.13138181E+02   # RIGHT SELECTRON
    35     2.13138181E+02   # RIGHT SMUON
    36     2.12838756E+02   # RIGHT STAU
    41     4.74810720E+02   # LEFT 1ST GEN. SQUARKS
    42     4.74810720E+02   # LEFT 2ND GEN. SQUARKS
    43     4.20057207E+02   # LEFT 3RD GEN. SQUARKS
    44     4.61771920E+02   # RIGHT U-SQUARKS
    45     4.61771920E+02   # RIGHT C-SQUARKS
    46     3.36293030E+02   # RIGHT T-SQUARKS
    47     4.60437692E+02   # RIGHT D-SQUARKS
    48     4.60437692E+02   # RIGHT S-SQUARKS
    49     4.60051305E+02   # RIGHT B-SQUARKS
    61     2.00000000E-01   # LAMBDA
    62     1.47434810E-01   # KAPPA
    63    -7.50350192E+01   # A_LAMBDA
    64    -2.18338774E+00   # A_KAPPA
# 
\end{verbatim}}
\end{table}

\begin{table}[p]
{\footnotesize
\baselineskip 10pt
\begin{verbatim}
BLOCK MASS   # Mass spectrum 
#  PDG Code     mass             particle 
        25     9.74781199E+01   # lightest neutral scalar
        35     4.48583968E+02   # second neutral scalar
        45     5.16024735E+02   # third neutral scalar
        36     2.19434909E+01   # lightest pseudoscalar
        46     4.50932452E+02   # second pseudoscalar
        37     4.55175450E+02   # charged Higgs
   1000001     4.96428647E+02   #  ~d_L
   2000001     4.80320535E+02   #  ~d_R
   1000002     4.91385003E+02   #  ~u_L
   2000002     4.80115856E+02   #  ~u_R
   1000003     4.96428647E+02   #  ~s_L
   2000003     4.80320535E+02   #  ~s_R
   1000004     4.91385003E+02   #  ~c_L
   2000004     4.80115856E+02   #  ~c_R
   1000005     4.40278168E+02   #  ~b_1
   2000005     4.77873225E+02   #  ~b_2
   1000006     3.03200260E+02   #  ~t_1
   2000006     5.24782130E+02   #  ~t_2
   1000011     2.43747516E+02   #  ~e_L
   2000011     2.16456923E+02   #  ~e_R
   1000012     2.33286301E+02   #  ~nue_L
   1000013     2.43747516E+02   #  ~mu_L
   2000013     2.16456923E+02   #  ~mu_R
   1000014     2.33286301E+02   #  ~numu_L
   1000015     2.15119676E+02   #  ~tau_1
   2000015     2.44554112E+02   #  ~tau_2
   1000016     2.33153313E+02   #  ~nutau_L
   1000021     5.01506590E+02   #  ~g
   1000022     7.55014082E+01   # neutralino(1)
   1000023     1.38801726E+02   # neutralino(2)
   1000025    -3.58593816E+02   # neutralino(3)
   1000035     3.80957538E+02   # neutralino(4)
   1000045     5.24690943E+02   # neutralino(5)
   1000024     1.37945343E+02   # chargino(1)
   1000037     3.79628706E+02   # chargino(2)
# Low energy observables
BLOCK LOWEN
         1     3.33858054E+00   # BR(b -> s gamma)*10^4
# 3*3 Higgs mixing
BLOCK NMHMIX
  1  1    -3.33228276E-01   # S_(1,1)
  1  2    -9.41055786E-01   # S_(1,2)
  1  3     5.80768815E-02   # S_(1,3)
  2  1    -9.29046853E-01   # S_(2,1)
  2  2     3.17227652E-01   # S_(2,2)
  2  3    -1.90364284E-01   # S_(2,3)
  3  1    -1.60719818E-01   # S_(3,1)
  3  2     1.17390906E-01   # S_(3,2)
  3  3     9.79994140E-01   # S_(3,3)
\end{verbatim}}
\end{table}

\begin{table}[p]
{\footnotesize
\baselineskip 10pt
\begin{verbatim}
# 3*3 Pseudoscalar Higgs mixing
BLOCK NMAMIX
  1  1     3.16227766E-01   # P_(1,1)
  1  2    -9.48683298E-01   # P_(1,2)
  1  3     0.00000000E+00   # P_(1,3)
  2  1    -9.84143132E-02   # P_(2,1)
  2  2    -3.28047711E-02   # P_(2,2)
  2  3    -9.94604680E-01   # P_(2,3)
  3  1     9.43564848E-01   # P_(3,1)
  3  2     3.14521616E-01   # P_(3,2)
  3  3    -1.03737795E-01   # P_(3,3)
# 3rd generation sfermion mixing
BLOCK STOPMIX  # Stop mixing matrix
  1  1     5.65184261E-01   # Rst_(1,1)
  1  2     8.24964697E-01   # Rst_(1,2)
  2  1    -8.24964697E-01   # Rst_(2,1)
  2  2     5.65184261E-01   # Rst_(2,2)
BLOCK SBOTMIX  # Sbottom mixing matrix
  1  1     9.94871315E-01   # Rsb_(1,1)
  1  2     1.01148731E-01   # Rsb_(1,2)
  2  1    -1.01148731E-01   # Rsb_(2,1)
  2  2     9.94871315E-01   # Rsb_(2,2)
BLOCK STAUMIX  # Stau mixing matrix
  1  1     1.82923365E-01   # Rsl_(1,1)
  1  2     9.83127175E-01   # Rsl_(1,2)
  2  1    -9.83127175E-01   # Rsl_(2,1)
  2  2     1.82923365E-01   # Rsl_(2,2)
# Gaugino-Higgsino mixing
BLOCK NMNMIX  # 5*5 Neutralino Mixing Matrix
  1  1     9.76939885E-01   # N_(1,1)
  1  2    -1.21822079E-01   # N_(1,2)
  1  3     1.55584479E-01   # N_(1,3)
  1  4    -8.03150977E-02   # N_(1,4)
  1  5     9.52878280E-03   # N_(1,5)
  2  1     1.69572219E-01   # N_(2,1)
  2  2     9.41177948E-01   # N_(2,2)
  2  3    -2.44926790E-01   # N_(2,3)
  2  4     1.58638593E-01   # N_(2,4)
  2  5    -1.65528778E-02   # N_(2,5)
  3  1    -4.43118962E-02   # N_(3,1)
  3  2     7.05108082E-02   # N_(3,2)
  3  3     6.98006382E-01   # N_(3,3)
  3  4     7.10367686E-01   # N_(3,4)
  3  5     3.50645553E-02   # N_(3,5)
  4  1    -1.21736484E-01   # N_(4,1)
  4  2     3.06778393E-01   # N_(4,2)
  4  3     6.47820106E-01   # N_(4,3)
  4  4    -6.79460614E-01   # N_(4,4)
  4  5     9.86388891E-02   # N_(4,5)
  5  1     7.09990264E-03   # N_(5,1)
  5  2    -1.60840169E-02   # N_(5,2)
  5  3    -9.44486010E-02   # N_(5,3)
  5  4     4.57635935E-02   # N_(5,4)
  5  5     9.94321905E-01   # N_(5,5)
\end{verbatim}}
\end{table}

\begin{table}[p]
{\footnotesize
\baselineskip 10pt
\begin{verbatim}
BLOCK UMIX  # Chargino U Mixing Matrix
  1  1     9.25480029E-01   # U_(1,1)
  1  2    -3.78796404E-01   # U_(1,2)
  2  1     3.78796404E-01   # U_(2,1)
  2  2     9.25480029E-01   # U_(2,2)
BLOCK VMIX  # Chargino V Mixing Matrix
  1  1     9.69142290E-01   # V_(1,1)
  1  2    -2.46501971E-01   # V_(1,2)
  2  1     2.46501971E-01   # V_(2,1)
  2  2     9.69142290E-01   # V_(2,2)
# 
# GAUGE AND YUKAWA COUPLINGS AT THE SUSY SCALE
BLOCK GAUGE Q=  4.68008177E+02 # (SUSY SCALE)
         1     3.60690129E-01   # g1(Q,DR_bar)
         2     6.45231920E-01   # g2(Q,DR_bar)
         3     1.10281535E+00   # g3(Q,DR_bar)
BLOCK YU Q=  4.68008177E+02 # (SUSY SCALE)
  3  3     9.40962309E-01   # HTOP(Q,DR_bar)
BLOCK YD Q=  4.68008177E+02 # (SUSY SCALE)
  3  3     4.53102195E-02   # HBOT(Q,DR_bar)
BLOCK YE Q=  4.68008177E+02 # (SUSY SCALE)
  3  3     3.17949988E-02   # HTAU(Q,DR_bar)
BLOCK L/K Q=  4.68008177E+02 # (SUSY SCALE)
         1     2.00784168E-01   # LAMBDA(Q,DR_bar)
         2     1.48000459E-01   # KAPPA(Q,DR_bar)
# 
# SOFT TRILINEAR COUPLINGS AT THE SUSY SCALE
BLOCK AU Q=  4.68008177E+02 # (SUSY SCALE)
  3  3    -3.96388267E+02   # ATOP
BLOCK AD Q=  4.68008177E+02 # (SUSY SCALE)
  3  3    -7.44522774E+02   # ABOT
BLOCK AE Q=  4.68008177E+02 # (SUSY SCALE)
  3  3    -3.22488355E+02   # ATAU
BLOCK AL/AK Q=  4.68008177E+02 # (SUSY SCALE)
         1    -8.42676828E+01   # ALAMBDA
         2    -2.56530300E+00   # AKAPPA
# 
# SOFT MASSES AT THE SUSY SCALE
BLOCK MSOFT Q=  4.68008177E+02 # (SUSY SCALE)
         1     8.16527293E+01   # M1
         2     1.53892994E+02   # M2
         3     4.72879088E+02   # M3
        21     5.81854464E+04   # M_HD^2
        22    -1.04142570E+05   # M_HU^2
        23    -1.32549686E+05   # M_S^2
        31     2.40063601E+02   # M_eL
        32     2.40063601E+02   # M_muL
        33     2.39934370E+02   # M_tauL
        34     2.13138181E+02   # M_eR
        35     2.13138181E+02   # M_muR
        36     2.12838756E+02   # M_tauR
        41     4.74810720E+02   # M_q1L
        42     4.74810720E+02   # M_q2L
        43     4.20057207E+02   # M_q3L
        44     4.61771920E+02   # M_uR
        45     4.61771920E+02   # M_cR
        46     3.36293030E+02   # M_tR
        47     4.60437692E+02   # M_dR
        48     4.60437692E+02   # M_sR
        49     4.60051305E+02   # M_bR
\end{verbatim}}
\end{table}

\begin{table}[p]
{\footnotesize
\baselineskip 10pt
\begin{verbatim}
#  MU_EFF:   3.53749043E+02 # (AT THE SUSY SCALE)
# 
# GAUGE AND YUKAWA COUPLINGS AT THE GUT SCALE
BLOCK GAUGEGUT MGUT=  2.39561249E+16 # (GUT SCALE)
         1     7.18934347E-01   # g1(MGUT,DR_bar), GUT normalization
         2     7.18934334E-01   # g2(MGUT,DR_bar)
         3     7.09148313E-01   # g3(MGUT,DR_bar)
BLOCK YUGUT MGUT=  2.39561249E+16 # (GUT SCALE)
  3  3     6.29869150E-01   # HTOP(MGUT,DR_bar)
BLOCK YDGUT MGUT=  2.39561249E+16 # (GUT SCALE)
  3  3     1.69658275E-02   # HBOT(MGUT,DR_bar)
BLOCK YEGUT MGUT=  2.39561249E+16 # (GUT SCALE)
  3  3     2.17473793E-02   # HTAU(MGUT,DR_bar)
BLOCK LGUT/KGUT MGUT=  2.39561249E+16 # (GUT SCALE)
         1     2.22229350E-01   # LAMBDA(MGUT,DR_bar)
         2     1.61430186E-01   # KAPPA(MGUT,DR_bar)
# 
# SOFT TRILINEAR COUPLINGS AT THE GUT SCALE
BLOCK AUGUT MGUT=  2.39561249E+16 # (GUT SCALE)
  3  3    -2.00014623E+02   # ATOP
BLOCK ADGUT MGUT=  2.39561249E+16 # (GUT SCALE)
  3  3    -1.99980396E+02   # ABOT
BLOCK AEGUT MGUT=  2.39561249E+16 # (GUT SCALE)
  3  3    -1.99994867E+02   # ATAU
BLOCK ALGUT/AKGUT MGUT=  2.39561249E+16 # (GUT SCALE)
         1    -1.99976861E+02   # ALAMBDA
         2    -2.29490634E+01   # AKAPPA
# 
# SOFT MASSES SQUARED AT THE GUT SCALE
BLOCK MSOFTGUT MGUT=  2.39561249E+16 # (GUT SCALE)
         1     2.00004409E+02   # M1
         2     2.00001910E+02   # M2
         3     1.99989075E+02   # M3
        21     3.99907796E+04   # M_HD
        22     4.00122196E+04   # M_HU
        23    -1.42907310E+05   # M_S
        31     3.99987978E+04   # M_eL
        32     3.99987978E+04   # M_muL
        33     3.99987958E+04   # M_tauL
        34     4.00004265E+04   # M_eR
        35     4.00004265E+04   # M_muR
        36     4.00004222E+04   # M_tauR
        41     3.99950362E+04   # M_q1L
        42     3.99950362E+04   # M_q2L
        43     3.99997864E+04   # M_q3L
        44     3.99948593E+04   # M_uR
        45     3.99948593E+04   # M_cR
        46     4.00046814E+04   # M_tR
        47     3.99956655E+04   # M_dR
        48     3.99956655E+04   # M_sR
        49     3.99956856E+04   # M_bR
\end{verbatim}}
\caption{Corresponding {\sf spectr.dat.test} output file.}
\label{slhaout}
\end{table}

\begin{table}[p]
{\footnotesize
\baselineskip 10pt
\begin{verbatim}
#
#  Total number of points scanned
#
1
#
#  Output format 0=short 1=long (not recommended for big scannings)
#
1
#
#  Computation of relic density using MicrOmegas (0=no, 1=yes)
#
1
#
#  lambda
#
.2D0
.2D0
#
#  kappa
#
.147D0
.147D0
#
#  tan(beta)
#
3.D0
3.D0
#
#  mu
#
353.D0
353.D0
#
#  A_lambda
#
-75.D0
-75.D0
#
#  A_kappa
#
-2.2D0
-2.2D0
#
#  Remaining soft terms (no scan)
#
mQ3=  420.D0
mU3=  336.3D0
mD3=  460.D0
mL3=  240.D0
mE3=  213.D0
AU3=  -396.4D0
AD3=  -744.5D0
AE3=  -322.5D0
mQ=   474.8D0
mU=   461.8D0
mD=   460.4D0
mL=   240.D0
mE=   213.1D0
M1=   81.7D0
M2=   153.9D0
M3=   472.9D0
\end{verbatim}}
\caption{The {\sf scaninp.dat} file for sample parameter scan.
\label{scaninpcase2}}
\end{table}

\end{document}